\documentclass{article}

\usepackage{PRIMEarxiv}

\usepackage[utf8]{inputenc} 
\usepackage[T1]{fontenc}    
\usepackage{hyperref}       
\usepackage{url}            
\usepackage{booktabs}       
\usepackage{amsfonts}       
\usepackage{nicefrac}       
\usepackage{microtype}      
\usepackage{lipsum}
\usepackage{amsmath}
\usepackage{fancyhdr}       
\usepackage{graphicx}       
\graphicspath{{media/}}     

\title{Link Node: A Method to Characterize the Chain Topology of Intrinsically Disordered Proteins

}

\author{
  \begin{minipage}[t]{0.3\textwidth}
    \centering
    Danqi Lang\\
    School of Physics \\
    Zhejiang University\\
    Hangzhou, 310058 \\
    PR China \\
    \texttt{danqi.lang@ens.psl.eu}
  \end{minipage}%
  \hfill
  \begin{minipage}[t]{0.3\textwidth}
    \centering
    Le Chen \\
    School of Physics \\
    Zhejiang University\\
    Hangzhou, 310058 \\
    PR China \\
    \texttt{12245050@zju.edu.cn}
  \end{minipage}%
  \hfill
  \begin{minipage}[t]{0.3\textwidth}
    \centering
    Jingyuan Li \\
    School of Physics \\
    Zhejiang University\\
    Hangzhou, 310058 \\
    PR China \\
    \texttt{jingyuanli@zju.edu.cn}
  \end{minipage}%
}

\begin{document}
\maketitle

\begin{abstract}
    
    It is well-known that intrinsically disordered proteins (IDP) are highly dynamic, which is related to their functionality in various biological processes. However, the characterization of the intricate structures of IDP has been a challenge. Here, we analyze the chain topology of IDPs to characterize their conformations, in combination with molecular dynamics simulation (MD). We systematically compute the Gauss Linking Number ($GLN$) between segments in IDP, and show that the resulting GLN Map can effectively depict an unconventional structure -- physical link, i.e., the entanglement between two segments. The crossing points of physical links are further identified and denoted as Link Nodes. We show that the probability distribution of Link Nodes is highly heterogeneous and there are certain residues that largely affect the chain topology of IDP. Moreover, the structural fluctuations of the vicinity of these residues are largely suppressed, i.e., Link Node provides useful information about the topological constraint imposed on the residues during the conformation fluctuations of IDP. We further reveal that the evolution of the chain topology is considerably slow (with a timescale of hundreds of nanoseconds), which is distinct from the flipping of residue contact.

\end{abstract}


\section{Introduction}

    Intrinsically disordered proteins (IDPs) are involved in various biological functions such as signal transduction, gene regulation and cell differentiation. \cite{dyson2005intrinsically,bondos2021roles} IDPs are highly dynamic, \cite{bujnicki2008prediction, uversky2000natively} which are believed to be related to their functionality. \cite{dunker2005flexible, kim2008role} Moreover, the conformational space of IDPs is highly complex. \cite{uversky2013unusual} While some indicators such as the radius of gyration (\textit{Rg}) \cite{lobanov2008radius, 2020Computing} and asphericity (\textit{Asphe}) \cite{rudnick1986aspherity, arkin2013gyration} are commonly used to describe the IDP conformation, they fail to provide a comprehensive depiction of IDP structures. For example, Figure \ref{S1} shows two conformations of a typical IDP -- the RGG domain\cite{ozdilek2017intrinsically} of LAF-1\cite{elbaum2015disordered}. Although the values of \textit{Rg} and \textit{Asphe} are comparable between these two conformations (with \textit{Rg} being 2.61 nm and \textit{Asphe} being 0.07 and 0.08, respectively), their structures are apparently different. Specifically, one conformation exhibits compaction in the C-terminal region (Figure \ref{S1}a, left panel), while the other exhibits compaction in the middle region (Figure \ref{S1}b, left panel).  Accordingly,  \textit{Rg} and \textit{Asphe} are not sufficient to fully characterize the structures of IDPs.

    As revealed in previous studies, topology analysis is an effective approach to characterize the unconventional structures of proteins. \cite{dabrowski2017topological} This method aims to characterize the topological structures of peptide chain, \cite{dabrowski2017tie} such as links, \cite{yan2001design,boutz2007discovery,dabrowski2016linkprot}knots, \cite{1994Are,lua2006statistics,virnau2006intricate,yeates2007knotted,jamroz2015knotprot} and lassos \cite{haglund2014pierced,2016Complex,2016LassoProt} as well as related pseudo-topological structures, \cite{taylor2001protein} like physical links \cite{caraglio2017physical} and pseudo knots \cite{staple2005pseudoknots}. The depiction of chain topology is conceived as a supplement of the canonical methods (e.g. secondary structures \cite{sun2004overview}). It should be noted that the structures of IDPs are irregular with the absence of secondary structures. Therefore, it is interesting to characterize the structure of IDP with regard to its chain topology.

    Here we proposed to analyze the physical link within IDP chain, i.e. the entanglement between two segments in IDP chain. More specifically, the segments of IDP chain are regarded as directed sub-chains, and their intertwining is estimated by the classical Gauss Linking Number (GLN) method. The GLN method can depict the relative spatial relationship between two directed sub-chains and measures how one sub-chain winds around the other. \cite{banchoff1976self} The GLN of these two standard links (formed by two closed arcs) are +1 and -1 (according to their relative direction, Figure \ref{S2}). \cite{kusner1997distortion} Since the sub-chains of IDP have open tails, they cannot form standard link: they intertwine with each other instead (i.e. forming physical link, Figure \ref{fig1}b \& \ref{fig1}c). The corresponding GLN falls within the range of (-1, 1). \cite{2020-GLN} On the basis of the GLN method, the relative spatial relationship of various pairs of sub-chains throughout the whole IDP chain is then systematically studied. And the resulting GLN Map can effectively depict the topology of IDP chain and illustrate all physical links. The crucial residues of the physical link, the site where two sub-chains intersect, are then identified and denoted as the Link Node. The IDP conformation and its evolution are then characterized with regard to the identified link node. 

    In this work, we use this method to characterize the conformation of a typical IDP -- the LAF-1 RGG domain. Its conformations are obtained from all-atom molecular simulation. We compute the GLN Maps of the LAF-1 RGG, as well as the corresponding Link Nodes. We noticed the profile of Link Node probability is highly heterogeneous and there are certain residues that largely affect the topology of the whole IDP chain. We further analyzed the conformation evolution of IDP on the basis of GLN Maps. Notably, the timescale of the evolution can reach hundreds of nanoseconds: the chain topology of IDP is considerably stable which is distinct from the flipping of residue contact. In summary, GLN Map with Link Node can serve as an effective method to depict the IDP conformation and its evolution.

\section{Models and Methods} 
\subsection{Physical Link}

    The topology analysis of the protein chain has been primarily focused on knots or links connected by covalent bonds. However, in the case of  IDP, the chain is open and lacks covalent cross-linking. Therefore, the canonical topology analysis of IDP is significantly limited. Pseudo-topology is then utilized. \cite{taylor2001protein} If the covalent bonding criterion is relaxed, i.e., artificially joining the two ends of the protein chain to form a circle, the chain topology IDP can be analyzed. Physical link (or probabilistic link \cite{10.1093/nar/gkw976} ) is a typical structure within the scope of pseudo-topology, which is comprised of two open arcs that cross each other. \cite{caraglio2017physical} By suitably closing each arc,  \cite{sumners1990detecting,millett2005linear,2011Probing,sulkowska2013knotting,millett2013identifying} the resulting structure is a standard link (i.e. Hopf. Link, Figure \ref{fig1}c). The orientation of the physical link is consistent with that of the corresponding standard link, either positive or negative (Figure \ref{S2}). 

\subsection{GLN Algorithm}

    We utilize the $GLN$ algorithm to characterize physical links within IDP chains. $GLN$ comes from the Clugreanu-White self-linking formula, \cite{10.2307/2373348} and it can depict the spatial relationship of two intertwined curves on the basis of the number of windings and its sign reflected the orientation. The definition is as follows:
    
     \begin{equation}
        \begin{aligned} GLN\equiv \frac{1}{4\pi }\oint _{\gamma _1}\oint _{\gamma _2} \frac{\vec {r}^{(1)}-\vec {r}^{(2)}}{|\vec {r}^{(1)}-\vec {r}^{(2)}|^3}\cdot \big (d\vec {r}^{(1)}\times d\vec {r}^{(2)}\big )
        \end{aligned}
    \label{eq1}
    \end{equation} where $\vec {r}^{(1)}$ and $\vec {r}^{(2)}$ are the spatial coordinates of two curves. For proteins, curves are the collections of positions of C-$\alpha$ atoms in two sub-chains (denoted by Chain A and Chain B), and the integral can be replaced by sums. \cite{banchoff1976self}
   
    \begin{equation}
        \begin{aligned} GLN\equiv \frac{1}{4\pi }\sum _{i=1}^{N_1-1}\sum _{j=1}^{N_2-1}\frac{\vec {R}_{i}^{(A)} -\vec {R}_j^{(B)}}{|\vec {R}_i^{(A)}-\vec {R}_j^{(B)}|^3}\cdot \left( d\vec {R}_i^{(A)}\times d\vec {R}_j^{(B)}\right) 
        \end{aligned}
    \end{equation}
    
    where ${R}_i^{(k)}=({r}_{i+1}^{(k)}+{r_i}^{(k)})/2$; and $r_i$ represents the coordinates of the $C_{\alpha}$ atoms in residues i of Chain A. Since physical links are not standard links, the $GLN$ of physical link is not an integer and it falls within the range of (-1, 1). And its absolute value can be served as the potential of linking. $GLN$ = 0 indicates the absence of a physical link.

\subsection{The sequence length of sub-chains}

\begin{figure}[htbp]
            \centering
            \includegraphics[width = 0.9\textwidth]{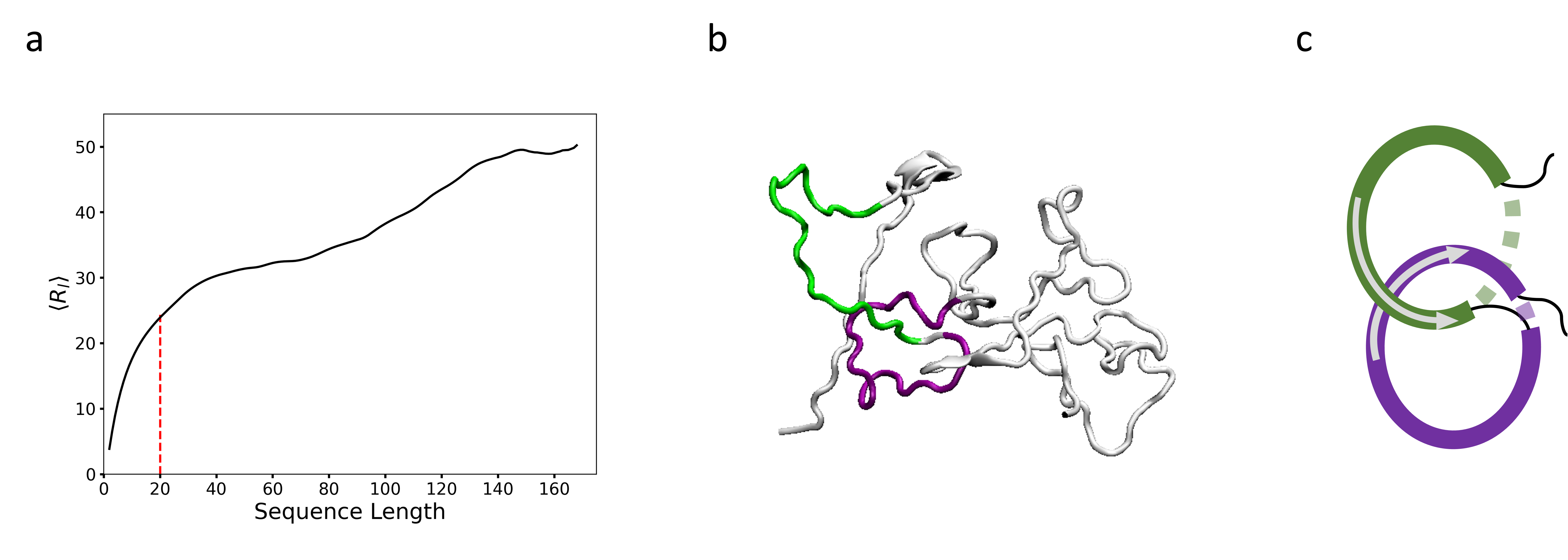}
            \caption{ The setup of sequence length in LAF-1 RGG domain and the corresponding physical link. (a) Average inter-residue distance $\langle R_l \rangle$ profiles for sequence length $l$. $\langle R_l \rangle$ reaches a plateau at approximately $l$ = 20 (red dashed line). (b) The snapshot of a representative conformation with a typical physical link consisting of Chain A (green, $l$ = 20) and Chain B (purple, $l$ = 20). (c) The corresponding schematic picture.}
            \label{fig1}
        \end{figure}

    The calculation of $GLN$ between various sub-chains within IDP requires the setup of the length for sub-chain. We calculate the average inter-residue distance $\langle R_l \rangle$ with varying sequence lengths of  $l$. \cite{2013Conformations} Figure \ref{fig1} presents the $\langle R_l \rangle$ profile of a representative LAF-1 RGG conformation. Clearly, $\langle R_l \rangle$ reaches a plateau at approximately $l=20$ (Figure \ref{fig1}a). And Figure \ref{fig1}b represents a snapshot of two sub-chains (with $l=20$) crossing each other in arc shapes. Their intertwining can be regarded as a standard Hopf. Link under a moderate closure way (Figure \ref{fig1}c). The GLN of this topological structure is $\sim {0.5}$, and it can be regarded as a typical physical link.

\subsection{GLN Map}

    We use $GLN_{i,j}$  to represent the $GLN$ between Chain A [i, i+l] and Chain B [ j, j+l]. There are 168 residues in LAF-1 RGG domain. The GLNs between various sub-chains are studied systematically using the Python package Topoly (version 0.9.17) \cite{2020-topoly}. The resulting 127$\times$127 matrix is defined as GLN Map (denoted as $M$):

    \begin{equation}
        M \equiv
        \left(\begin{array}{cccc}
                GLN_{1,22} & GLN_{1,23} & \cdots & GLN_{1,148}\\
               &GLN_{2,23} & \cdots &GLN_{2,148} \\
               & & \ddots & \vdots \\
               & &&GLN_{127,148}
            \end{array}\right)
    \end{equation}

\subsection{Trajectory preparing the structures of LAF-1 RGG protein}
    The procedure of the simulation of LAF is described in our previous study. \cite{zhang2022sequence} In order to mimic physiological conditions, sodium and chloride ions are added to neutralize the system, resulting in a 150 mM NaCl concentration. Five independent 1000-ns simulations to are conducted to characterize the structure of IDPs. 

\section{Results}
\subsection{The GLN Map of the LAF-1 RGG Domain}

\begin{figure}[htbp]
        \centering
        \includegraphics[width = 0.9\textwidth]{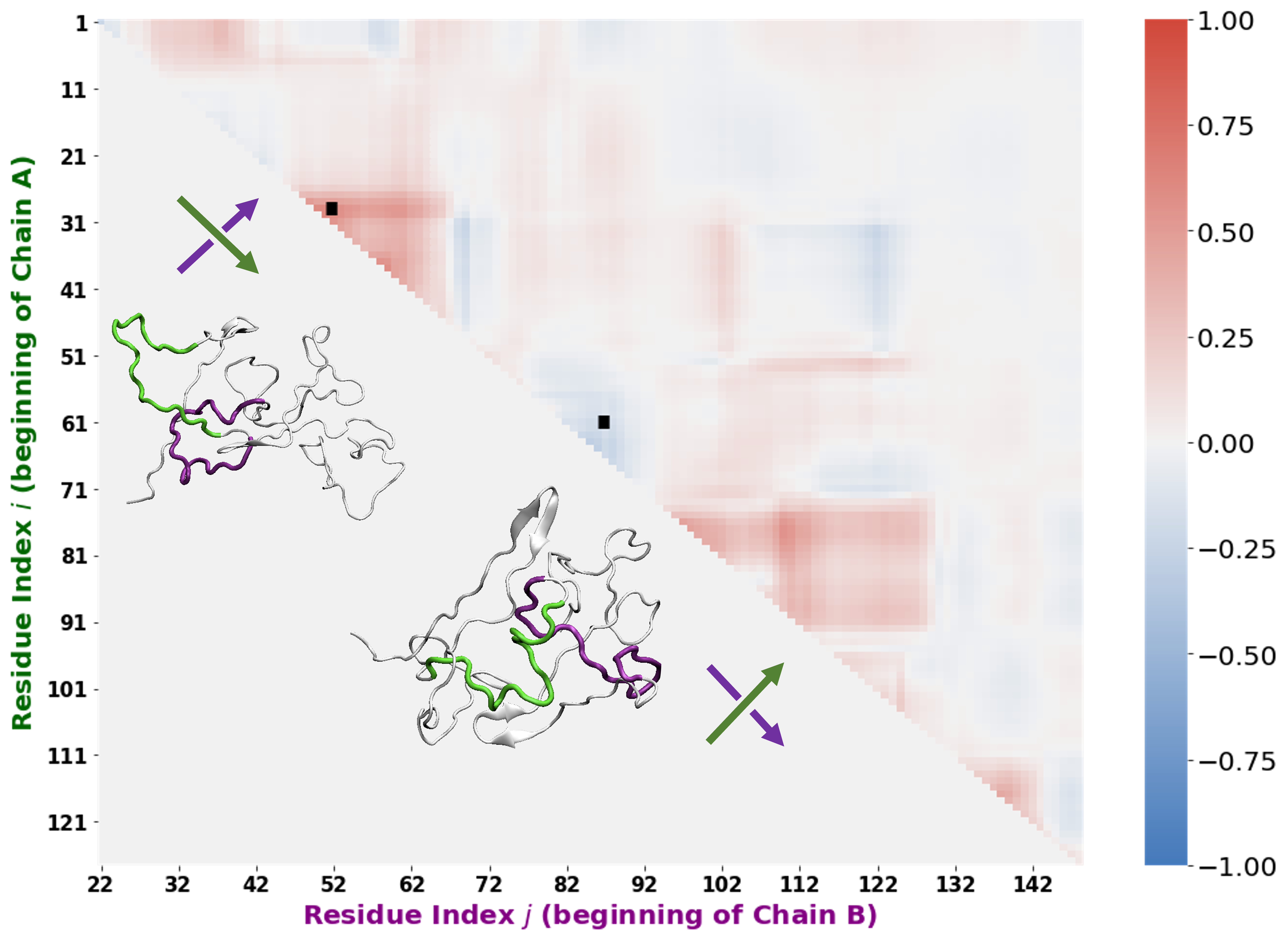}
        \caption{The GLN Map of a representative conformation. The x-axis represents the beginning residue of Chain A, and the y-axis represents the beginning residue of Chain B. Each cell on the GLN Map represents the GLN of Chain A and Chain B. Red and blue indicate the positive and negative directions respectively. The corresponding snapshots of the marked cells are represented nearby, where Chain A is highlighted in green and Chain B is in purple.}
        \label{fig2}
    \end{figure}

    We introduce the GLN Map method to characterize the chain topology of LAF-1 RGG domain. Sub-chain containing 20 residues is considered, and the GLNs of various pairs of these sub-chains (denoted by Chain A and Chain B, separately) are calculated (See the Materials and Methods section for details). One representative GLN Map and the corresponding conformation are shown in Figure \ref{fig2}. It can be seen that there are several patches in the GLN Map, especially near the diagonal (marked by black squares), which correspond to the physical links within IDP chain (see the snapshots in Figure \ref{fig2}). Notably, there is an alternating color pattern on the diagonal (i.e. three red-blue-red patches) are three apparent patches. This pattern illustrates that there are multiple crossings of sub-chains and the presence of multiple physical links in IDP conformation. Thus, the chain topology of IDP can be effectively depicted in terms of the arrangement of crossings (i.e. physical links).

    It should be noted that the GLN Map can effectively discriminate the IDP conformations with similar \textit{Rg} and \textit{Asphe}. Figure \ref{S1}a \& \ref{S1}b show the GLN Maps of the two IDP conformations with comparable \textit{Rg} ($\sim 2.61$ nm) and \textit{Asphe} ($\sim 0.07$). The patterns of these two GLN Maps are apparently different. As for the conformation in Figure \ref{S1}a, the corresponding GLN Map shows a typical patch in its upper left corner (corresponding to the N-terminal part of RGG domain), indicating the presence of a physical link in this region. On the other hand, there is no patch in this region of Figure \ref{S1}b. Hence, the chain topology of these two IDPs is different, even though they share similar profiles.

    Moreover, the evolution of IDP conformation can be analyzed with regard to the change in chain topology on the basis of the trajectory of  GLN Maps. Figure \ref{S4} shows GLN Maps at t = 600ns,750ns, and 900ns of a representative trajectory. At t = 600ns, there is a physical link between sub-chains 61-81 and 87-107; at t = 750ns, these two sub-chains become parallel and the physical link is released; at t = 900ns, they form a physical link again but with the reverse orientation. In other words, their spatial relationship is reversed. The conformation change can be attributed to the rearrangement of chain topology of IDP. Hence, the analysis of chain topology may provide useful insight into the conformation evolution of IDP.

\subsection{Identification of the Link Node in RGG Domain}

    In the GLN map, there are emergence of color patches, i.e. the regions with a considerable color gradient. This alternation of chain topology can be quantitively reflected in the $GLN$ changing from zero to nonzero. Specifically, the change of the amplitude of  $GLN$ is calculated and denoted as i.e. $\Delta \lvert GLN \rvert$:
    \begin{equation}
       \Delta \left|GLN_{i,j}^{h}\right| =  \frac{\lvert GLN_{i,j+1}\rvert-\lvert GLN_{i,j-1}\rvert}{\lvert GLN_{i,j}\rvert}
   \end{equation}
   \begin{equation}
       \Delta \left|GLN_{i,j}^{v} \right|= \frac{\lvert GLN_{i+1,j} \rvert - \lvert GLN_{i-1,j}\rvert }{\lvert GLN_{i,j}\rvert} 
   \end{equation}Where $\Delta \lvert GLN \rvert^{h}$ and $\Delta \lvert GLN \rvert^{v}$represent the change of $\lvert GLN \rvert^{h}$ in the horizontal and vertical axis of GLN Map, respectively.

   $\Delta \lvert GLN \rvert$ is then exploited to identify the boundary of patch. Firstly, the sites with large $\Delta \lvert GLN \rvert$ are identified, according to the criterion $\Delta \lvert GLN \rvert >$ 50\%. Secondly, those consecutive sites are chosen and the corresponding line of these sites is then identified as the boundary of patch. Boundaries with a length beyond 20 (the plateau of $\langle R_l \rangle$ profile is around 20, Figure \ref{fig1}a) are considered in the following discussion. Let’s take the red patch in Figure \ref{fig2} (marked by a black square) as an example. The sites with $\lvert GLN \rvert^{v}$  exceeding 50\% are identified (see the enlarged GLN Map in Figure \ref{S6}). These sites are consecutive and located at line i = 27 (Figure \ref{S6}). Thus i = 27 can be considered as a horizontal boundary of this color patch. And the length of this boundary line is beyond 20. Similarly, the identified vertical boundary is j = 67 (Figure \ref{S7}).

   The residues in these boundaries are further discussed on the basis of the sign of $\Delta \lvert GLN \rvert$. If $\Delta \lvert GLN \rvert >$ 0, the position of boundary line corresponds to the e terminal residue of this sub-chain is about to cross with another sub-chain. For instance, line i = 27 indicates the terminal residue of sub-chain 27-47 is about to cross with the other sub-chain (Figure \ref{S7}a). Likewise, if $\Delta \lvert GLN \rvert <$ 0, it indicates that the beginning residue of this sub-chain is about to cross with the other sub-chain. For instance, line j = 67 indicates the beginning residue of sub-chain 67-87 is about to cross with the other sub-chain (Figure \ref{S7}c). Therefore, these residues serve as the crossing points44 of the physical link and are then denoted as the Link Nodes (see the algorithm of Link Node identification in SI, Figure \ref{S5}).

    In summary, on the basis of the GLN Map, we analyze the boundary of the patch and identify the corresponding Link Node (e.g. residues 47 and 67). These Link Nodes are located at the sites where two sub-chains cross, and effectively depict the corresponding physical link. Hence, Link Nodes can be regarded as the pivotal residues to depict the chain topology of IDP.

\subsection{Characterization of IDP Structures on the Basis of Link Nodes}

\begin{figure}[htbp]
        \centering
        \includegraphics[width = 1\textwidth]{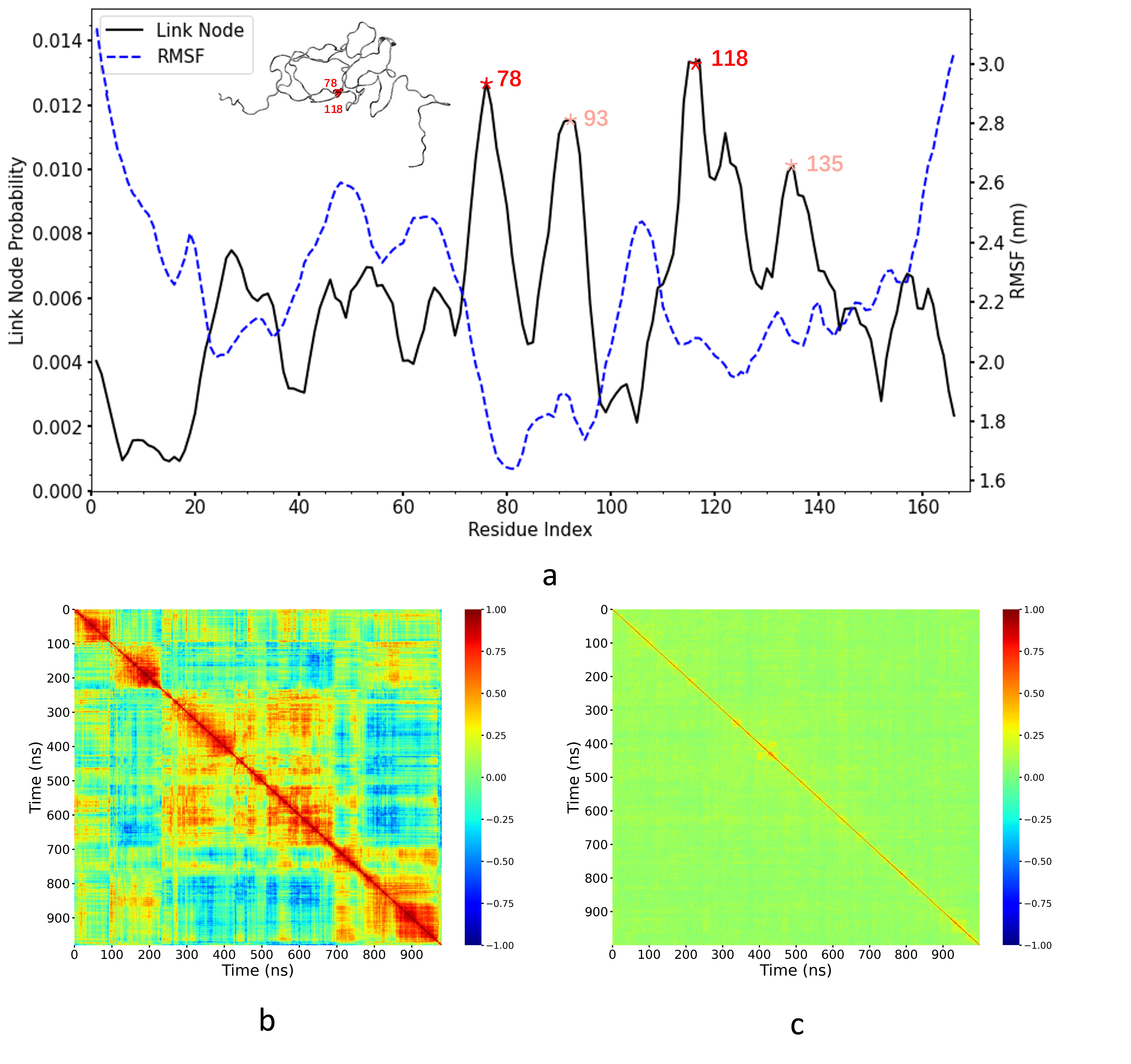}
        \caption{The Link Node probability distribution and the evolution of GLN Map. (a) The Link Node probability distribution (black line) and the RMSF (blue line). The snapshot of a representative conformation reveals that 78 and 118 are the crossing points. (b) The correlation coefficient matrix of GLN Maps. (c) The correlation coefficient matrix of contact maps.}
        \label{fig3}
    \end{figure}

    Link Nodes are then exploited to analyze the conformation evolution of LAF-1 RGG domain, and five independent 1000-ns trajectories are considered.  The Link Node probabilities of these five trajectories, i.e., the probability distributions of residues involved in Link Nodes, are calculated. Figure \ref{fig3}a shows the probability profile of one trajectory (the other four probability profiles are represented in Figure \ref{S11}). These probability profiles are highly heterogeneous (, even though the residues with higher propensity vary from trajectory to trajectory). As shown in Figure \ref{fig3}a, the propensity of some residues, i.e. 78Asp (1.20\%), 93Arg (1.15\%), 118Gly (1.34\%) and 135Asn (1.01\%), are about twice the average (0.58\%). A representative conformation illustrates the case 78Asp and 118Gly serving as Link Nodes (Figure \ref{fig3}a, upper left corner).

    To further illustrate the contribution of Link Node residues to the conformation evolution, the RMSF of each residue is calculated and compared with the Link Node probability.  The RMSF and Link Node probability exhibits an anti-correlation, with a Pearson correlation coefficient of $ -0.55$. Such a relationship is also observed in the other four trajectories, with correlation coefficients of $-0.56$, $-0.47$, $-0.40$, and $-0.51$, respectively (Figure \ref{S11}). Notably, the regions with high Link Node propensity (e.g. 78Asp, 93Arg, 118Gly and 135Asn) usually have low RMSF (Figure \ref{fig3}a). And the RMSFs of these residues are less than 2.0 nm, well below the average value (2.2 nm) In other words, the regions involved in Link Nodes tend to have suppressed structural fluctuation. Therefore, Link Node method should also provide useful information about the topological constraint imposed on the residues during the conformation fluctuation of IDP. 

    The conformation fluctuation of IDP is further analyzed from the perspective of chain topology. The evolution of chain topology is analyzed by calculating the correlation coefficient ($\rho$) among GLN Maps. $\rho$ is defined as follows:

    \begin{equation}
        \rho_{M(t_1),M(t_2)} = \frac{M(t_1)\odot M(t_2)}{\sqrt{\Vert M(t_1) \Vert \bullet \Vert M(t_2) \Vert}}
    \end{equation}

    where $M(t)$ are the GLN Maps at $t$. The resulting correlation matrix is shown in (Figure \ref{fig3}b). The correlation coefficient between the GLN Map with close instant is considerably high. And several blocks (with $\rho$ maintaining up to 0.75) are identified along the diagonal of the correlation matrix. Moreover, the time scale of these blocks generally reaches hundreds of nanoseconds. This suggests the evolution of the chain topology of IDP is quite slow. And there are sustained physical links which is involved in the conformation evolution of IDP. Figure \ref{S9} shows the GLN Maps at t = 310ns, 350ns and 390ns, which are in the same block in Figure \ref{fig3}b. Clearly, the physical link 70-90 and 110-130 is sustained during the conformation evolution This indicates that although the IDP undergoes significant conformational fluctuations, it still maintains similar chain topology. Additionally, the correlation matrices of the GLN Map of the other four trajectories are shown in Figure \ref{S12} and the time scale of the diagonal blocks also can reach hundreds of nanoseconds. A relatively slow evolution of the chain topology of IDP is also observed in these trajectories.

    The probability distribution of residue contacts is also calculated for comparison. The probability profile is rather homogeneous (Figure \ref{S8}). And there are much more residue contacts in IDP structures: there are $\sim$50 residue contacts in each conformation, while there are only $\sim$5 Link Nodes. In other words, the homogeneity of abundant residue contact makes it less effective to depict IDP structures. Besides, we also calculate the correlation coefficients between contact matrices to reveal the evolution of residue contacts (Figure \ref{fig3}c). Clearly, their correlation coefficients are generally low,  and there are no blocks can be identified on the correlation matrix. This suggests the evolution of residue contact is fast. And the residue contact cannot characterize the sustained feature during the conformation evolution of IDP.

\subsection{Mechanism underlying sustained chain topology}

\begin{figure}[htbp]
        \centering
        \includegraphics[width = 0.9\textwidth]{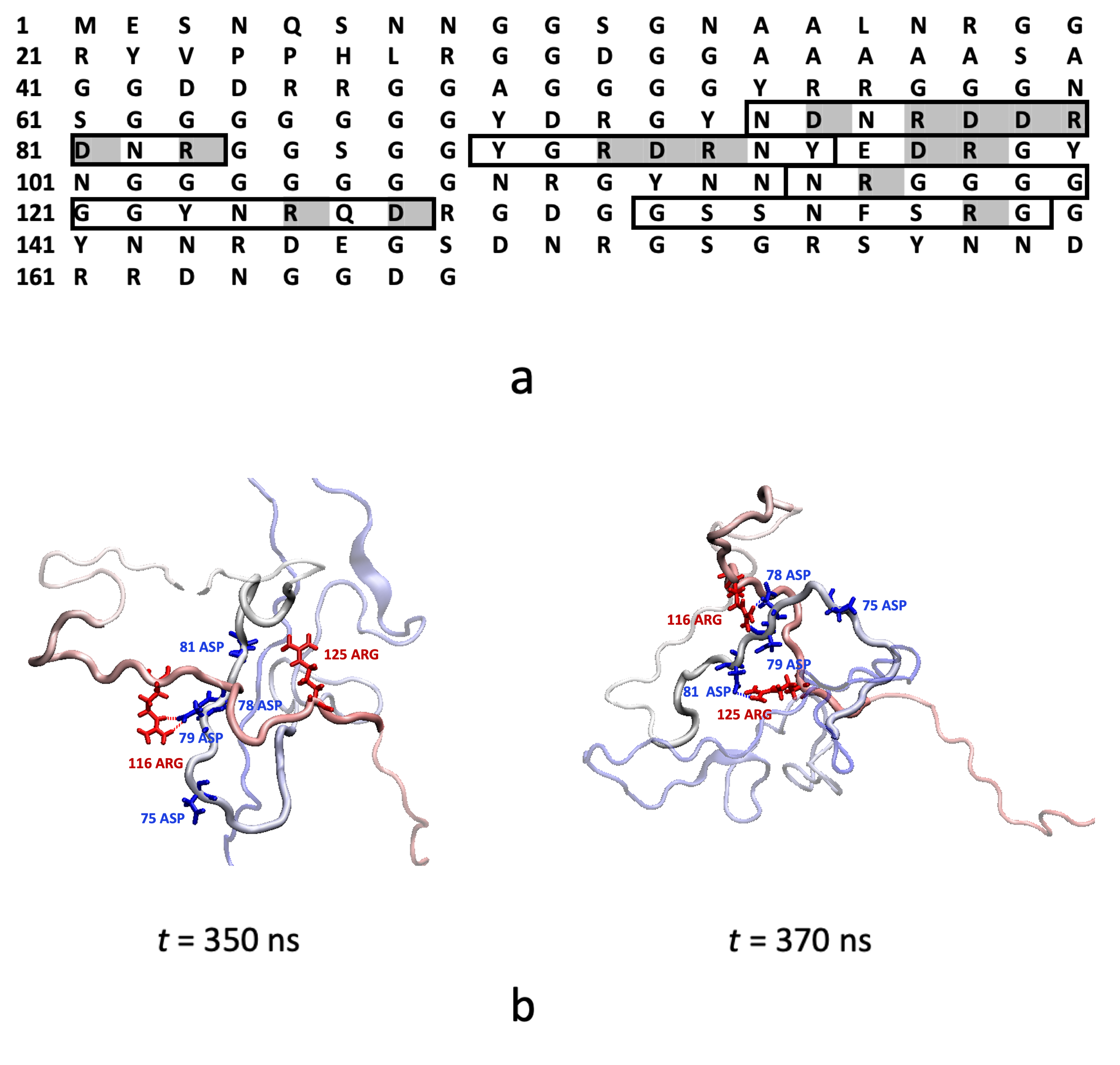}
        \caption{ 
        The interaction mode for the sustained physical link . (a) The sequence of the LAF-1 RGG domain. Four segments with higher Link Node propensity (74-83, 89-95, 115-127, 132-139) are outlined by rectangles. Charged residues are highlighted in grey. (b) Snapshots of representative sustained physical link. The IDP is colored in blue-to-red (from N-terminus to C-terminus) and segments 74-83 and 115-127 are highlighted in bold. Asp in red, Arg in blue.}
        \label{fig4}
    \end{figure}

    As revealed by our results about the correlation coefficient matrix of GLN map, the evolution of the chain topology of IDP is rather slow. To study the mechanism underlying such sustained chain topology, the interaction modes associated with physical links are then discussed. In the representative trajectory, the sustained physical links with the lifetime $>$50 ns are identified and the probability distribution of the Link Nodes of sustained physical links is shown in Figure \ref{S10}. Interestingly, the distribution of sustained Link Node is similar to the overall Link Nodes. The residues with higher Link Node propensity are further discussed,  i.e. larger than 1.0\%. There are 38 residues that can be categorized into four segments, i.e. 74-83, 89-95, 115-127, and 132-139, which are highlighted in grey in Figure \ref{fig4}a. We notice that there are a total of 14 charged residues (8 Arg and 6 Asp). The corresponding proportion of charged residues (37\%) is considerably higher than the overall proportion of the protein (26\%). Figure \ref{fig4}b shows a representative sustained physical link involving segments 74-83 and 115-127. There are four negatively charged residues (75Asp, 78Asp, 79Asp, and 81Asp) in segment 74-83, and two positively charged residues (116Arg and 125Arg) in segment 115-127. The interaction between two segments and the sustained physical link is the electrostatic attraction between these residues. At t = 350ns, the physical link is attributed to the interaction of 79Asp 116Arg; at t = 370ns, the physical link is attributed to both 78Asp-116Arg and 81Asp-125Arg. Overall, the electrostatic attraction mediates the formation and maintenance of such sustained physical link, even though the given interactions of charged residues are highly dynamic. 

\section{Conclusion}
    In this work, we analyze the chain topology of IDP to characterize its conformation. The GLN between the sub-chains is calculated throughout the IDP. And the resulting GLN Maps can identify the corresponding physical links which are comprised of two crossing sub-chains, thus effectively characterizing the chain topology of IDP. The crossing point of the physical link is identified according to the sub-chain orientation and denoted as Link Node. And the Link Node can serve as the key residue to depict the chain topology of IDP. We employ this Link Node method on the basis of the GLN Map to analyze the MD simulation trajectories of the LAF-1 RGG domain and identify the Link Nodes of all conformations. The probability profile of Link Node is highly heterogeneous, and the regions with higher propensity are abundant in charged residues such as Arg and Asp. The Link Nodes and the interaction between crossing sub-chains can be attributed to the electrostatic interaction between these residues. Moreover, the Link Node propensity is negatively correlated with RMSF, suggesting that the structural fluctuation of the given region is suppressed under the topological constraint. The conformational evolution of IDP is further depicted by analyzing the evolution of GLN map. The correlation coefficient matrix of GLN Map is calculated. Interestingly, the GLN map can remain considerably correlated over the timescale of hundreds of nanoseconds. Hence, the identified chain topology exhibits considerable dynamic stability. Taken together, the Link Node based on GLN Map is an effective method to characterize the conformations and conformational fluctuations of IDPs.

\section*{Acknowledgments}
    This work was supported by the National Natural Science Foundation of China (NSFC) (Grant Nos. 12175195 and 32371299).

\bibliographystyle{unsrt}  
\bibliography{references}  

\appendix
\setcounter{figure}{0}
\renewcommand{\thefigure}{S\arabic{figure}}

\newpage
\section{Appendixes}

\subsection{Conformations of similar \textit{Rg} and \textit{Asphe}}

\begin{figure}[htbp]
            \centering
            
            \includegraphics[width = 0.9\textwidth]{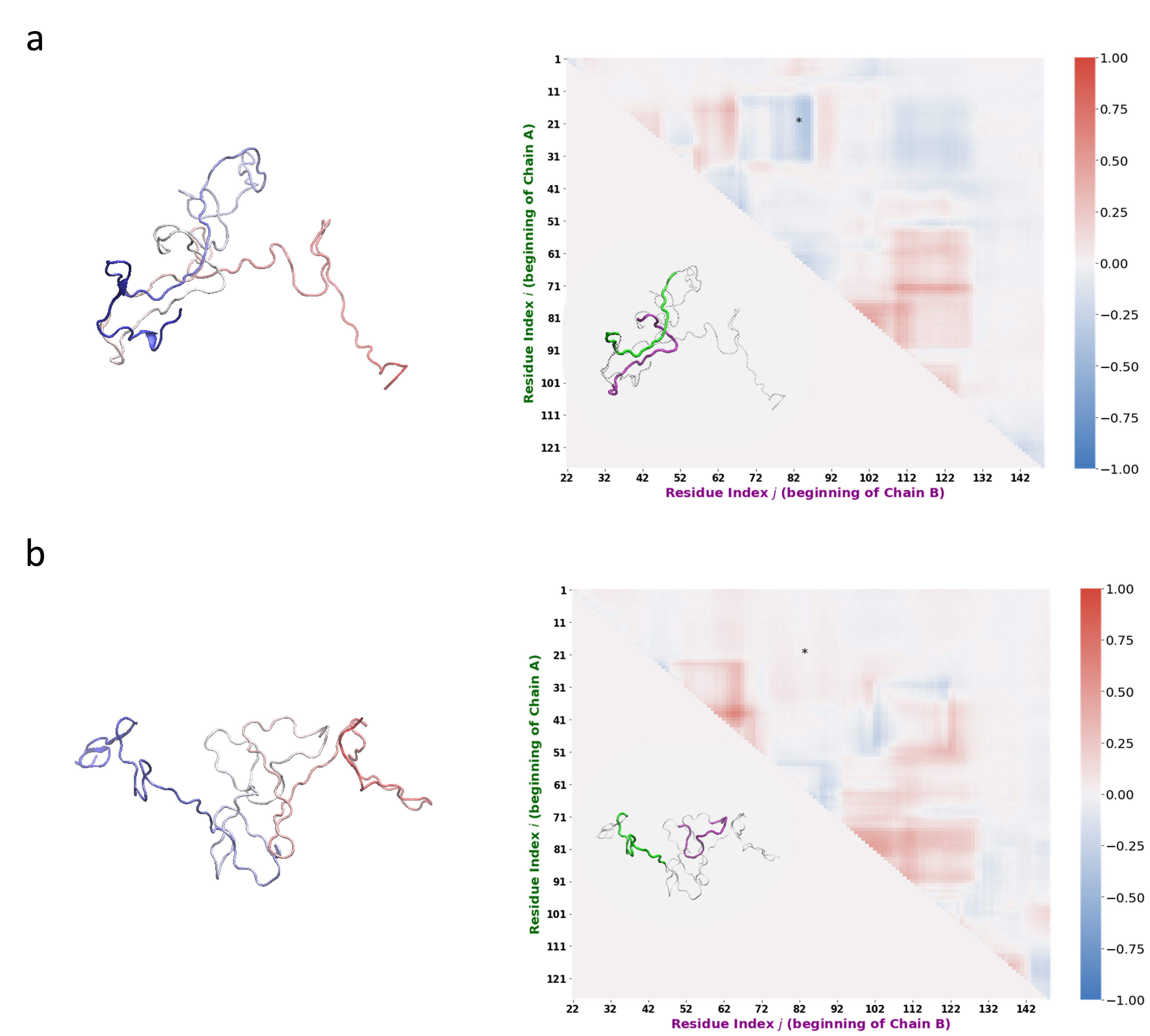}
            \caption{Conformations of similar \textit{Rg} and \textit{Asphe} with the corresponding GLN Maps. (a) \textit{Rg} = 2.61nm, \textit{Asphe} = 0.07; (b) \textit{Rg}= 2.61nm, \textit{Asphe} = 0.08.}
        \label{S1}
        \end{figure}

    Figure \ref{S1} shows two representative conformations with similar \textit{Rg} and \textit{Asphe}, while their GLN Maps are different.

\subsection{Gauss Linking Number ($GLN$)}

\begin{figure}[htbp]
            \centering
            
            \includegraphics[width = 0.9\textwidth]{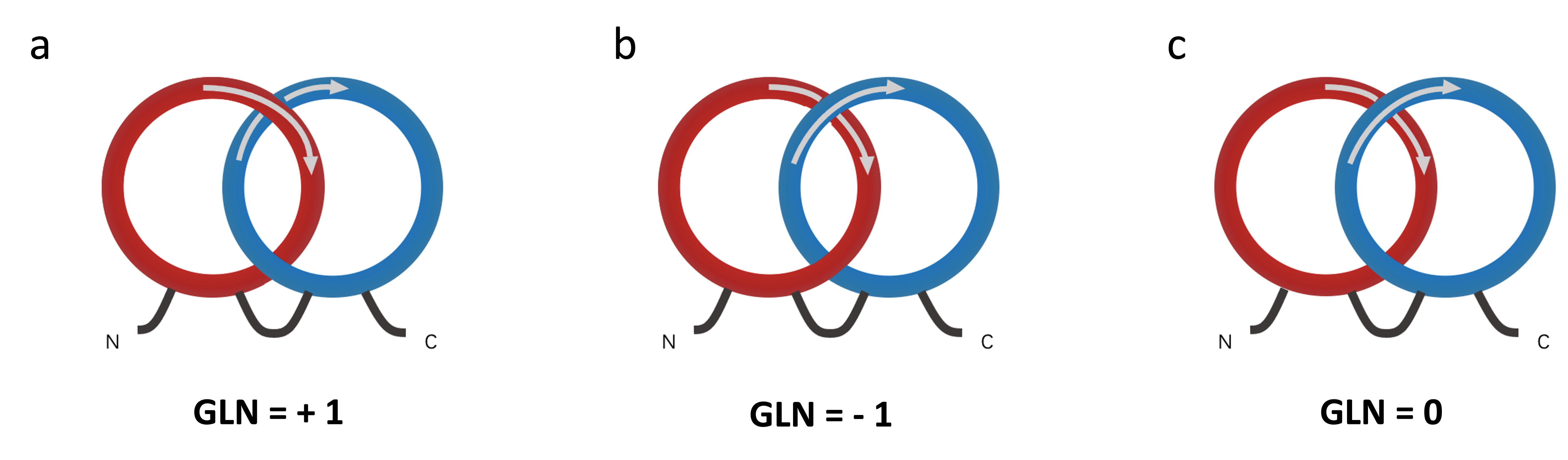}
            \caption{Gauss Linking Number ($GLN$). (a) Hopf.1 Link, $GLN $= +1, positive direction. (b) Hopf.2 Link,$GLN$ = -1, negative direction. (c) No link,$GLN$ = 0.}
        \label{S2}
        \end{figure}

    Figure \ref{S2} illustrates how to use Gauss Linking Number ($GLN$) to describe links.

\subsection{The color intensity of the GLN Map}

\begin{figure}[htbp]
            \centering
            
            \includegraphics[width = 0.9\textwidth]{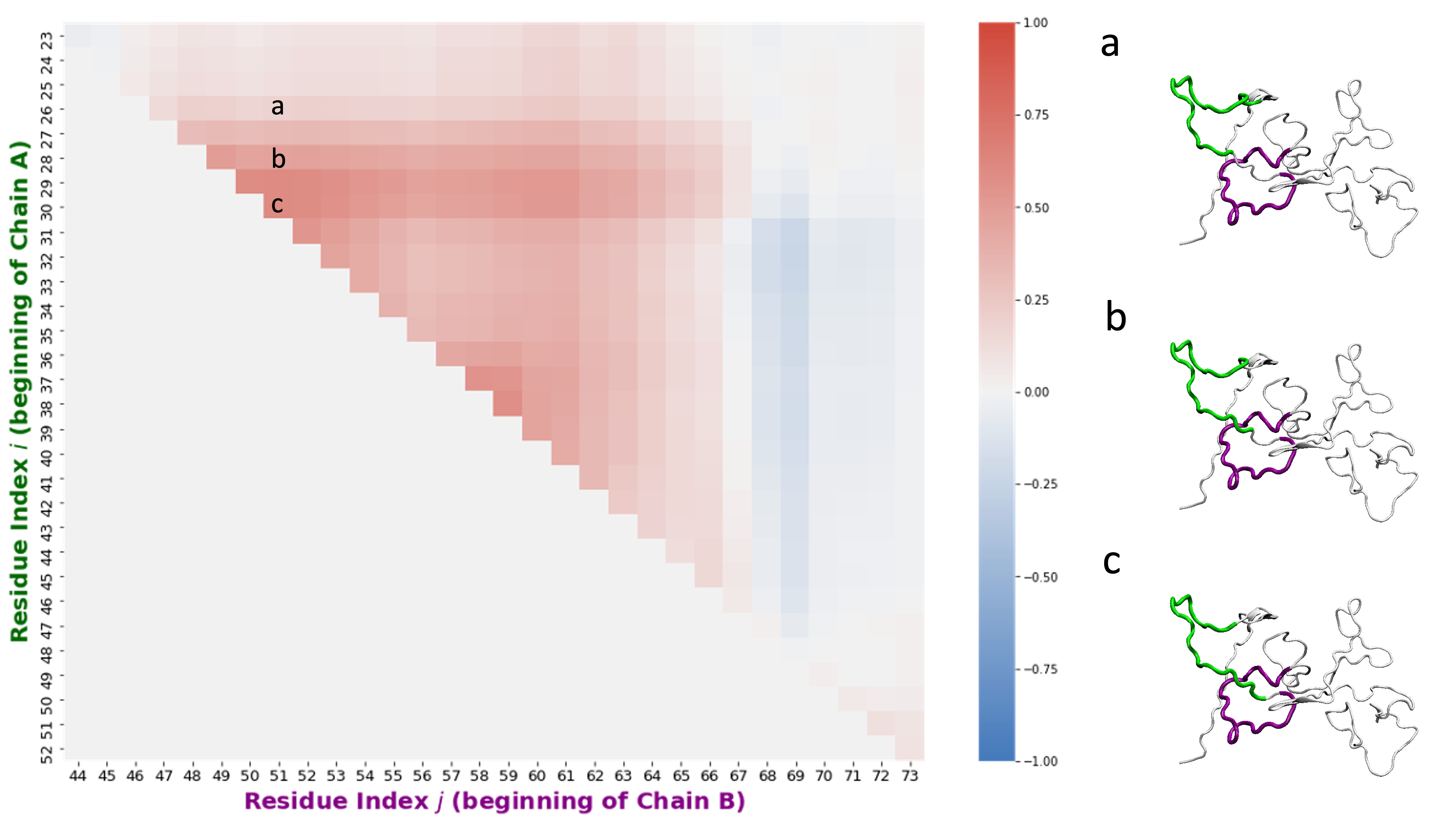}
            \caption{Interpretation of the color intensity of GLN Map. Left Panel: an inset of the GLN Map shown in Figure \ref{fig2}. Right panel: Corresponding Snapshots. Chain A is in green, Chain B is in purple. a: GLN = 0.189. b: GLN = 0.327. c: GLN = 0.496}
        \label{S3}
        \end{figure}

    Figure \ref{S3} is an inset of Figure \ref{fig2}. It can be seen that from site a to c (i.e., i varies from 26 to 30), Chain A (green, right panel) pokes deeper into the fixed Chain B(purple, right panel, with j = 51). This is revealed by the increase of the color intensity  (corresponding $GLN$ increases from 0.189 to 0.496). 
    
\subsection{Revealing the dynamic process by GLN Map}

    \begin{figure}[htbp]
            \centering
            
            \includegraphics[width = 0.5\textwidth]{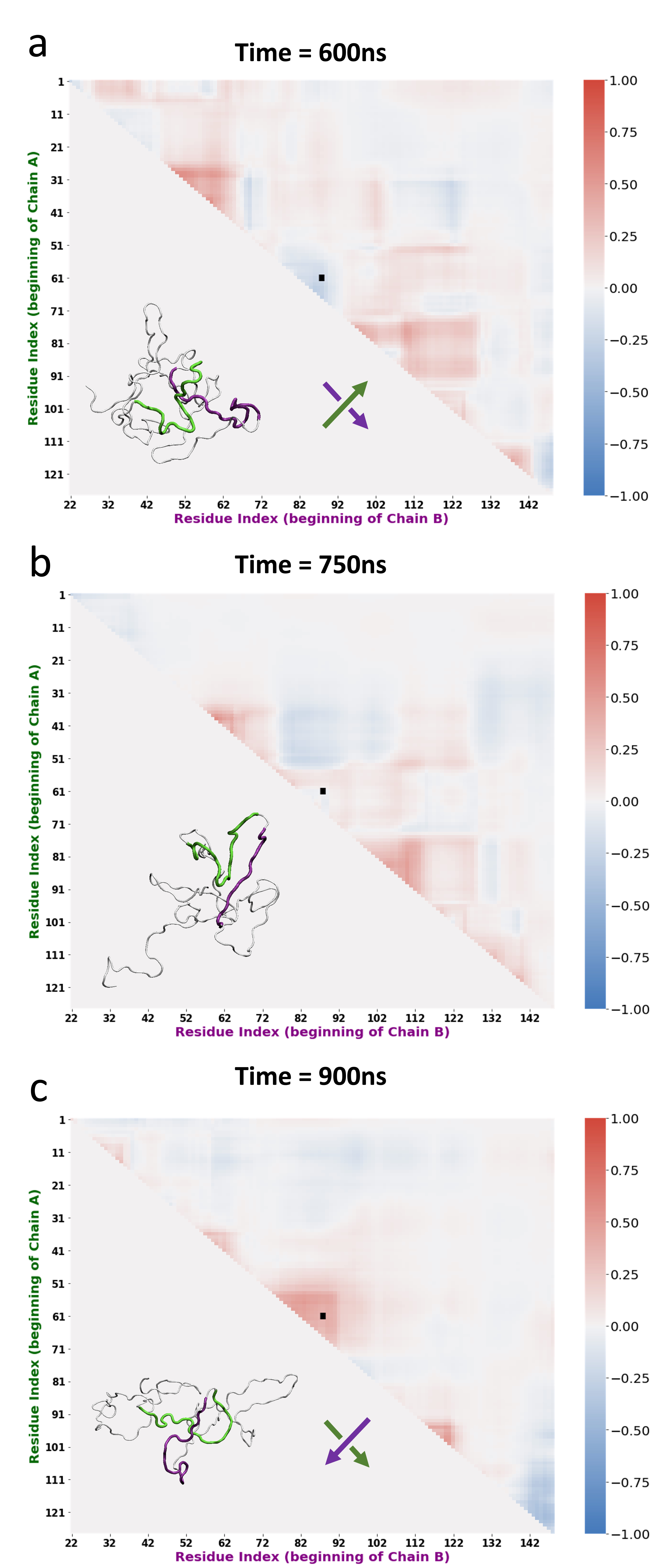}
            \caption{  GLN Map reveals the dynamic process of a typical physical link. (a) t = 600ns, form the physical link in the negative direction. (b) t = 750ns, the physical link is released. (c) t = 900ns, form the physical link in the positive direction.}
        \label{S4}
        \end{figure}

\newpage
\subsection{Identification of the Link Node}

\begin{figure}[htbp]
            \centering
            
            \includegraphics[width = 0.9\textwidth]{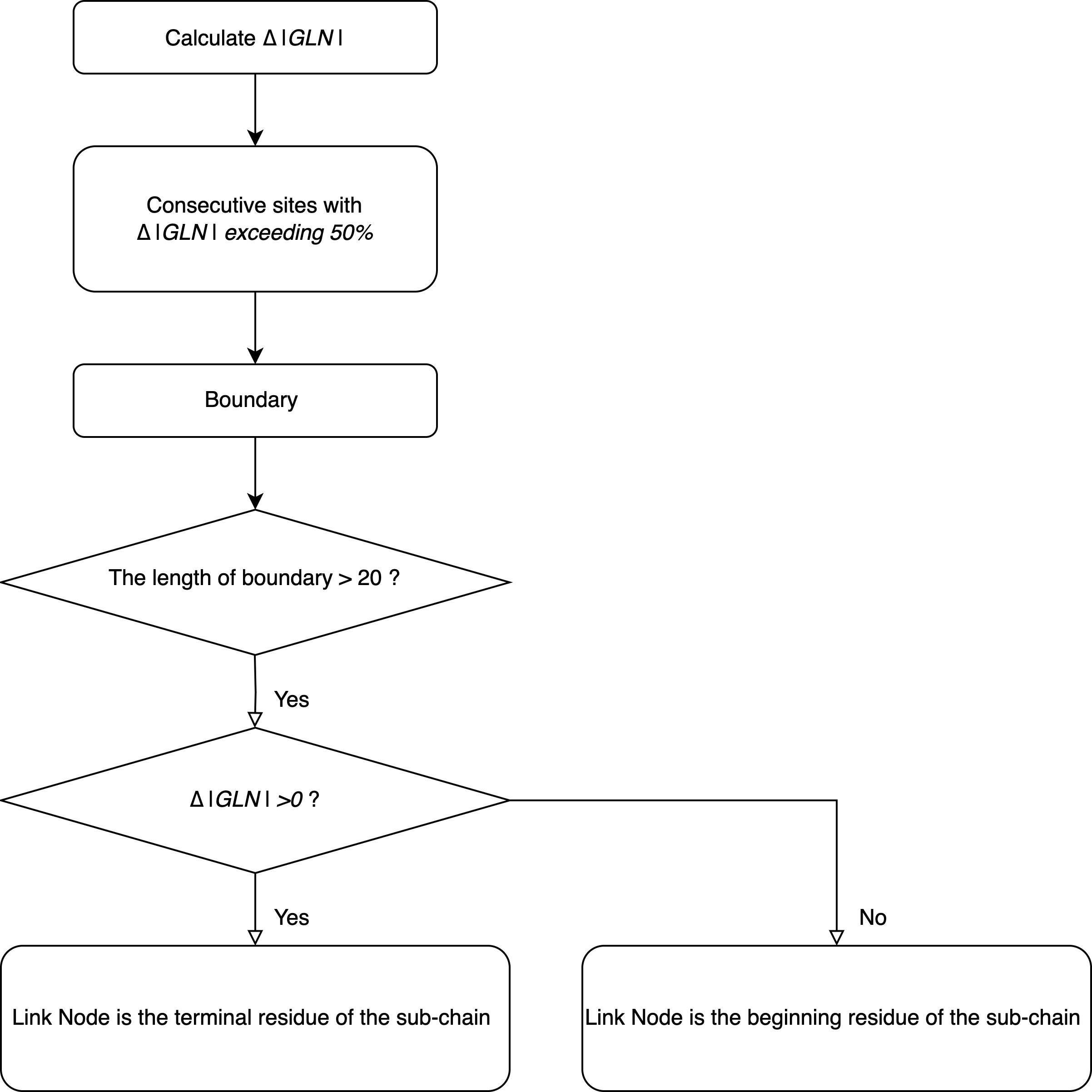}
            \caption{The process of Link Node identification.}
        \label{S5}
        \end{figure}

\begin{figure}[htbp]
            \centering
            
            \includegraphics[width = 0.9\textwidth]{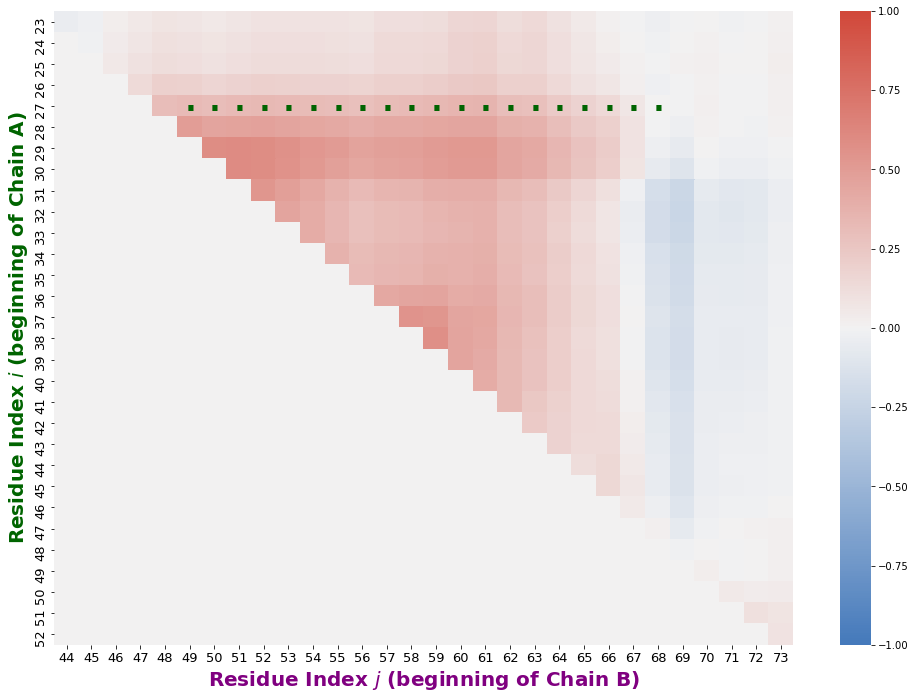}
            \caption{The variation of $GLN$. Square: Sites with $\Delta \lvert GLN \rvert $ exceed 50\%.}
        \label{S6}
        \end{figure}

    The process chart for Link Node identification is presented in Figure \ref{S5}. Firstly, we calculate $\Delta \lvert GLN \rvert $ to represent the variation of $\lvert GLN \rvert$. Then we identify sites with $\Delta \lvert GLN \rvert $ exceeding 50\% and the corresponding line is regarded as the boundary. We focus on the boundary with a length over 20 and Link Node is identified on the basis of the sign of $\Delta GLN$ of the corresponding boundary(Figure \ref{S6}).

\begin{figure}[htbp]
            \centering
            
            \includegraphics[width = 0.9\textwidth]{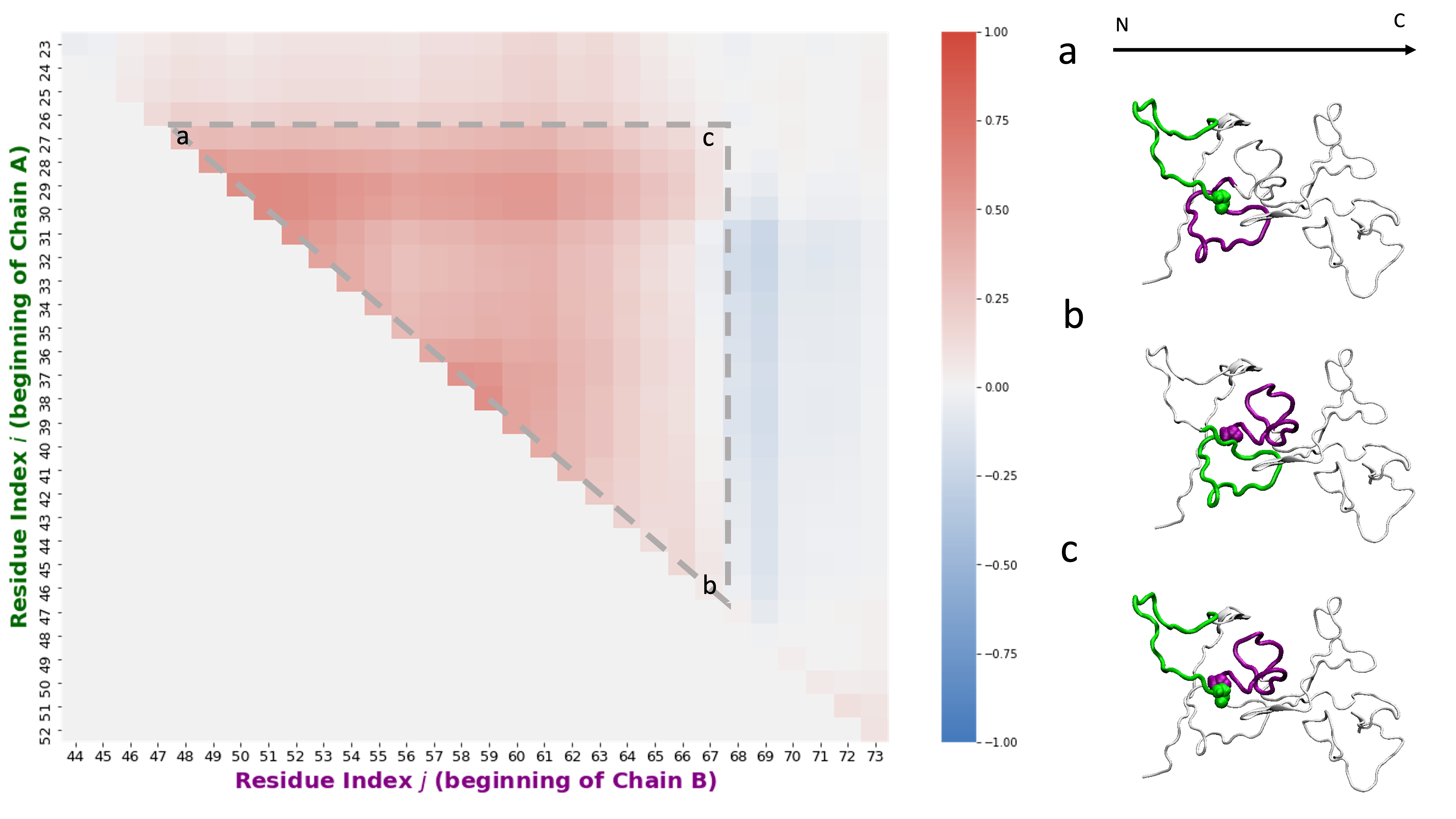}
            \caption{The boundary of a representative patch. Residue 47 and residue 67 are shown in VDW spheres (green:47 and purple:67). (a)The terminal residue of Chain A (47) is the crossing point. (b) The beginning residue of Chain B (67) is the crossing point. (c) The crossing pair.}
        \label{S7}
        \end{figure}

    Accordingly, the boundaries of the red patch of Figure \ref{fig2} are shown in Figure \ref{S7}.  The line corresponding to the horizontal boundary (i = 27) and vertical boundary (j = 67) cross at site c, thus this patch can be delineated by the dashed gray line $\overline{ac}$ and $\overline{bc}$. For $\overline{ac}$, the corresponding $\lvert GLN \rvert > $ 0, indicates the presence of a physical link between Chain A and Chain B in this region. Within a modestly long variation of Chain B (i.e. j ranges from 48 to 68), the terminal residue of Chain A, Residue 47 (i.e., i = 27, plus the sequence length 20, the terminal residue is 47) is crossing through Chain B (Figure \ref{S7}a and Figure \ref{S7}c). As Chain A moves towards the C-terminus, the crossing between Chain A and Chain B becomes deeper, which is reflected as $\Delta \lvert GLN \rvert > $ 0 on $\overline{ac}$. On the other hand, for bc, the $\lvert GLN \rvert$> 0, Figure \ref{S7}b and \ref{S7}c display that the beginning residue of Chain B, Residue 67, crosses through Chain A and thus forms the physical link. If Chain B continues to move towards the C-terminus, Chain A and Chain B will no longer cross, and physical links cannot exist anymore, which is reflected as $\Delta GLN$ <0 on $\overline{bc}$. Thus, these two crossing residues (47 and 67) are denoted as Link Nodes. Taken together, Link Node can be identified on the basis of the sub-chain corresponding to the boundary. If the $\Delta GLN$ of the boundary line is positive (reflected as color appears), Link Node is the terminal residue of the corresponding sub-chain; if the $ \Delta \lvert GLN \rvert $ of the boundary line is negative (reflected as color disappears), Link Node is the beginning residue of the corresponding sub-chain.
    
\subsection{Comparison of Link Node and contact method}

\begin{figure}[htbp]
            \centering
            
            \includegraphics[width = 0.9\textwidth]{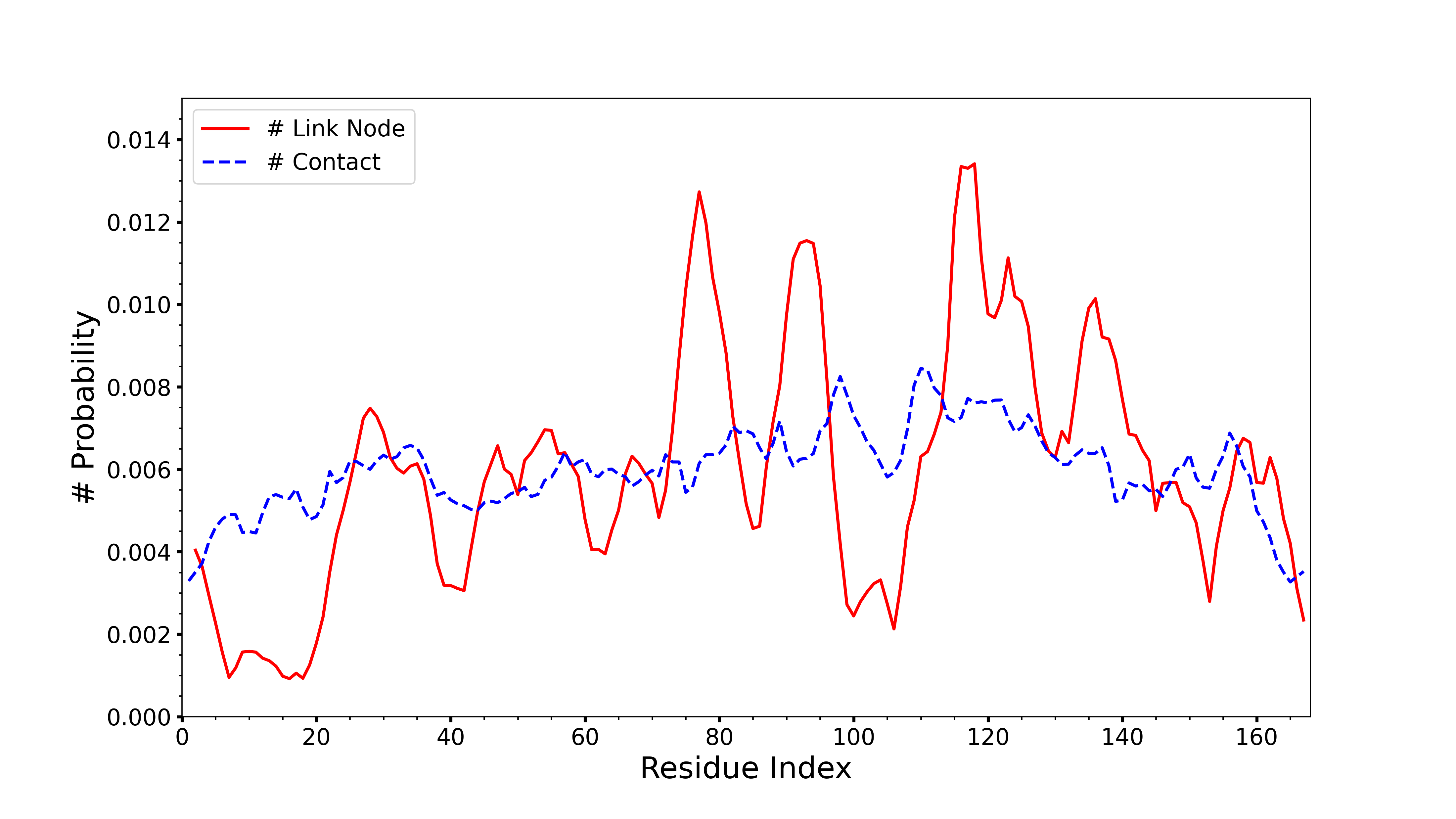}
            \caption{The Link Node (red) and contact (blue) probability of sequence profile.}
        \label{S8}
        \end{figure}

    Figure \ref{S8} shows the probability profile of Link Nodes and contacts.

\subsection{Sustained physical link}

\begin{figure}[htbp]
            \centering
            
            \includegraphics[width = 0.5\textwidth]{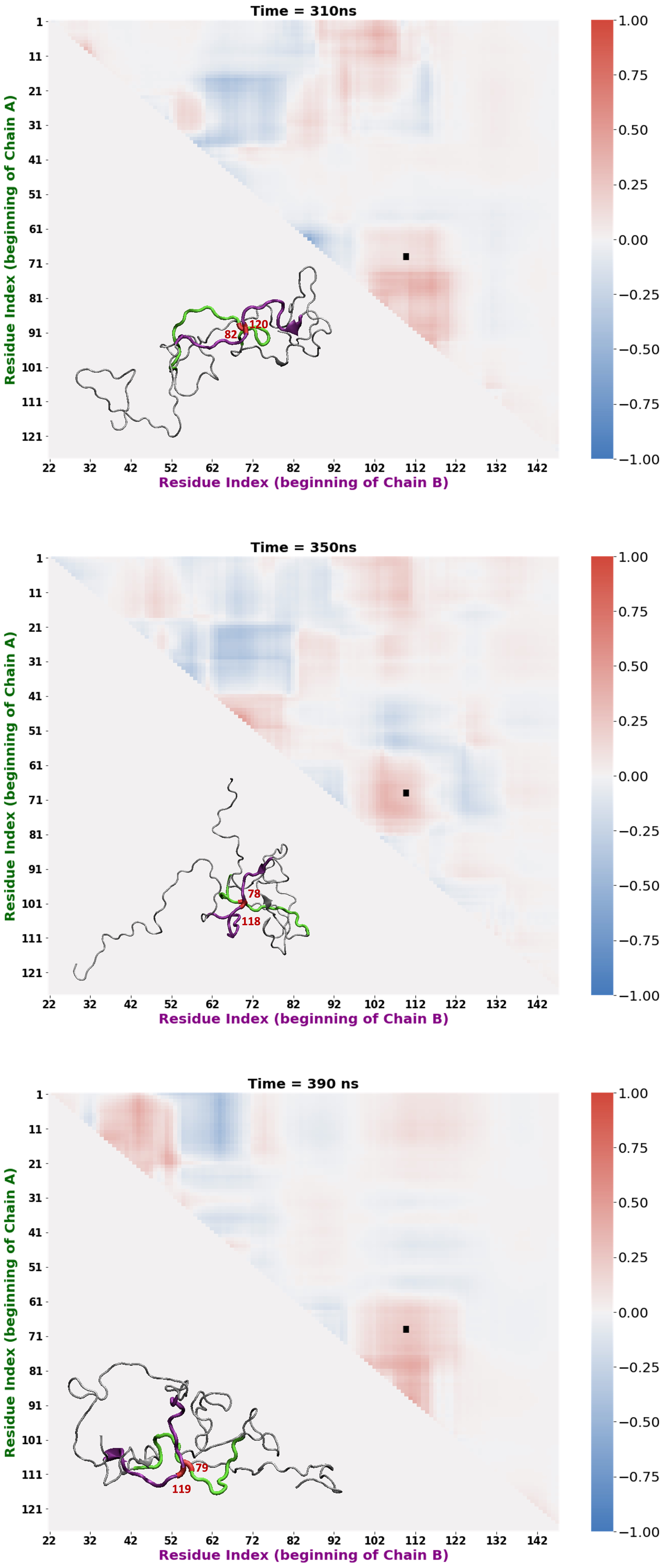}
            \caption{A sustained physical link. Dark square: 70-90 and 110-130.}
        \label{S9}
        \end{figure}

    Figure \ref{S9} shows a representative sustained physical link.


    \begin{figure}[htbp]
            \centering
            
            \includegraphics[width = 0.9\textwidth]{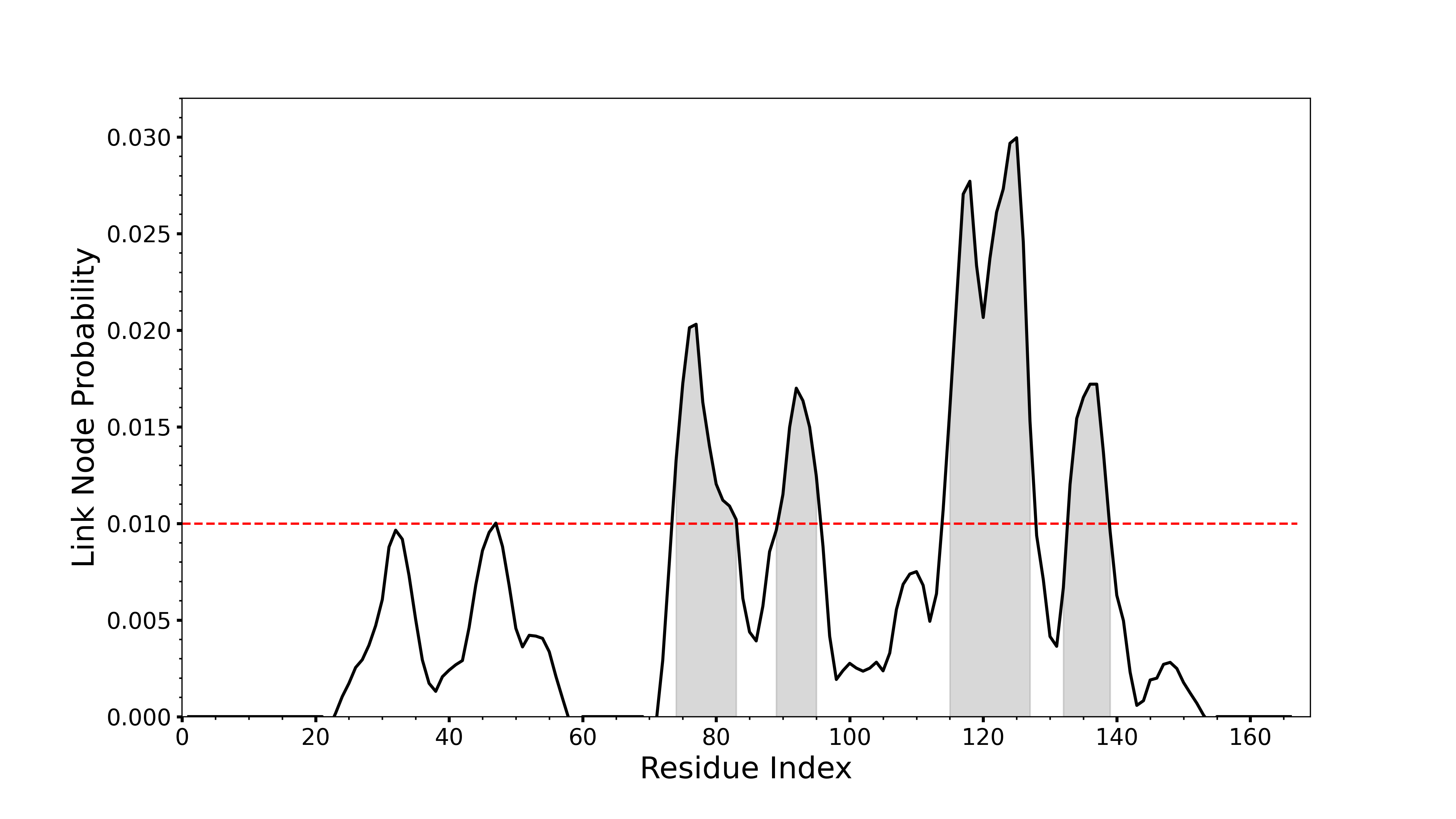}
            \caption{The Link Node probability distribution of sustained physical links.}
        \label{S10}
        \end{figure}

\subsection{Other trajectories}

    Figure \ref{S12} shows the probability profile of LInk Nodes in the other four trajectories.

        \begin{figure}[htbp]
            \centering
            
            \includegraphics[width = 1\textwidth]{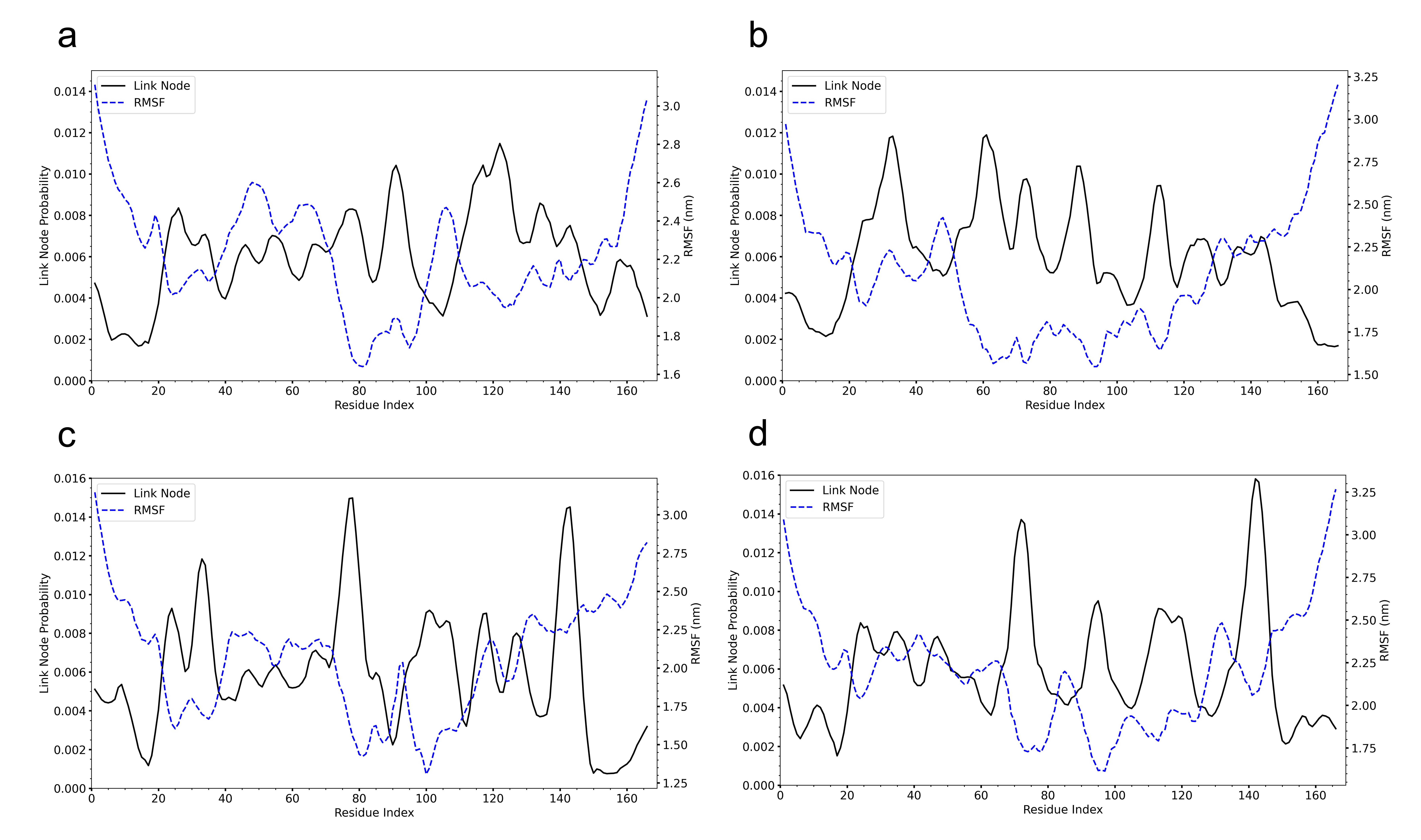}
            \caption{The Link Node probability of sequence profiles in other trajectories.}
        \label{S11}
        \end{figure}

        \begin{figure}[htbp]
            \centering
            
            \includegraphics[width = 0.9\textwidth]{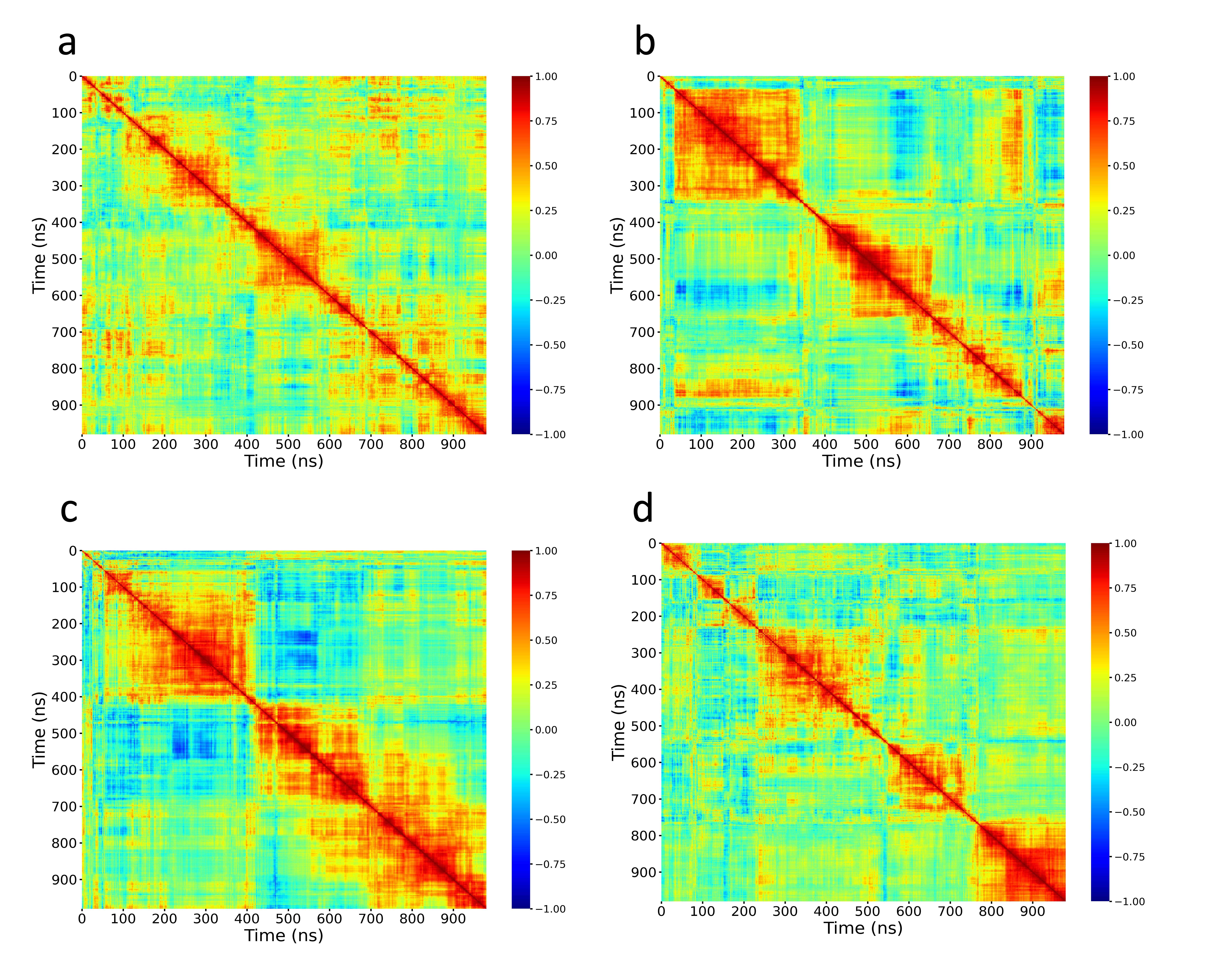}
            \caption{The correlation coefficient matrix of GLN Maps of conformations in other trajectories}
        \label{S12}
        \end{figure}

\subsection{Interaction Modes}

\begin{figure}[htbp]
            \centering
            
            \includegraphics[width = 0.9\textwidth]{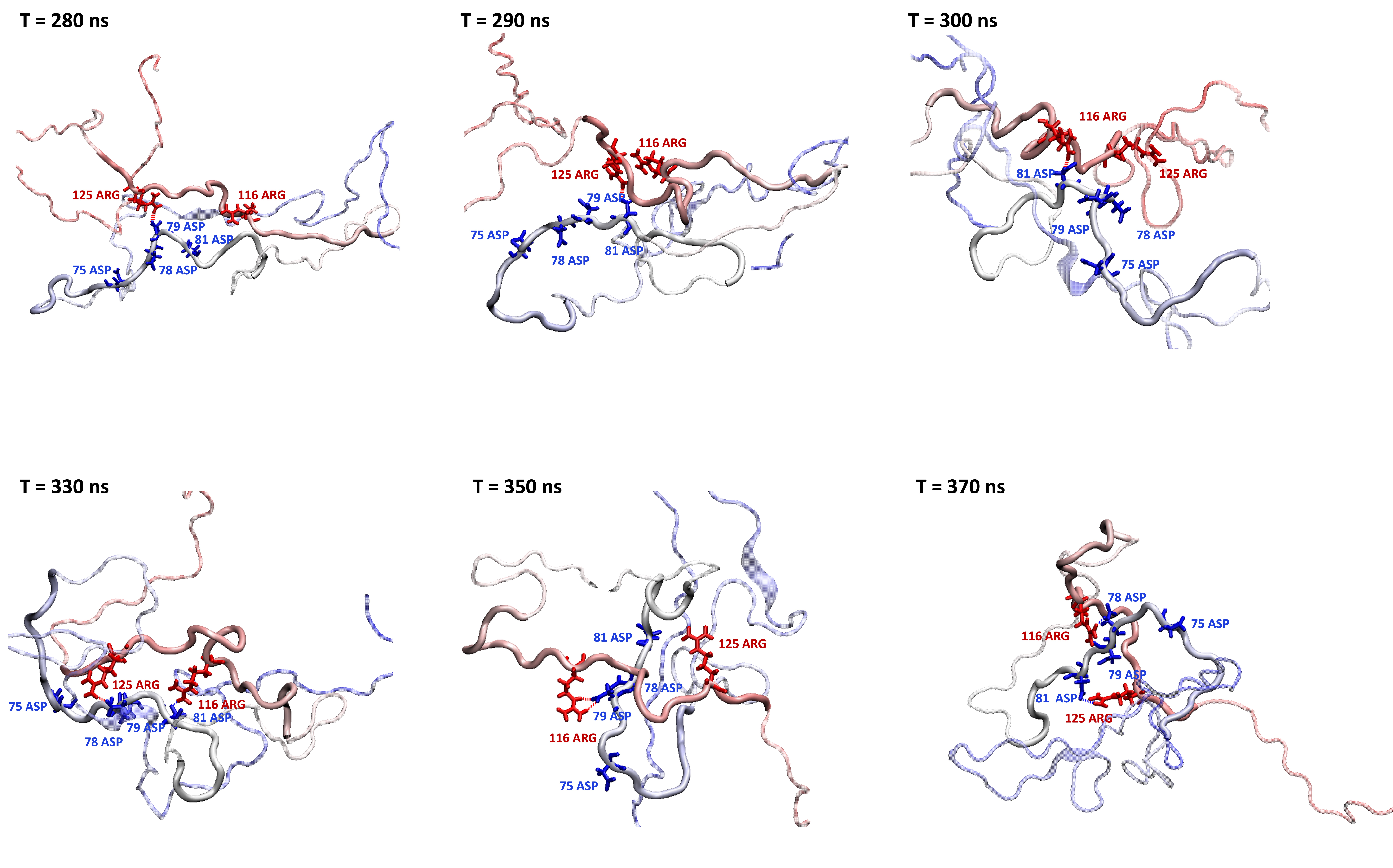}
            \caption{The interaction mode between ARG and ASP. The h-bond is in red and the salt bridge is in blue}
        \label{S13}
        \end{figure}

    Figure \ref{S13} shows an interaction mode between ARG and ASP.

\end{document}